\documentclass[english,prl,twocolumn, aps,amssymb,footinbib,showpacs]{revtex4-1}
\usepackage[T1]{fontenc}
\usepackage{amsmath}
\usepackage{amssymb}
\usepackage{graphicx}
\usepackage{braket}
\usepackage{babel}
\usepackage{color}

\makeatletter

\@ifundefined{textcolor}{}
{
 \definecolor{BLACK}{gray}{0}
 \definecolor{WHITE}{gray}{1}
 \definecolor{RED}{rgb}{1,0,0}
 \definecolor{GREEN}{rgb}{0,1,0}
 \definecolor{BLUE}{rgb}{0,0,1}
 \definecolor{CYAN}{cmyk}{1,0,0,0}
 \definecolor{MAGENTA}{cmyk}{0,1,0,0}
 \definecolor{YELLOW}{cmyk}{0,0,1,0}
 }

\usepackage{bm}\@ifundefined{definecolor}{\usepackage{color}}{}

\newcommand{\DD}[0]{\mathcal{D}}
\newcommand{\tr}[0]{\text{Tr}}
\newcommand{\bb}[0]{\hat{b}}

\newcommand{\bin}[0]{b_\mathrm{in}}

\usepackage[T1]{fontenc}
\usepackage[latin9]{inputenc}
\usepackage{textcomp}
\usepackage{color}

\usepackage{babel}

\makeatother

\begin{document}

\title{Detecting itinerant microwave photons with engineered non-linear dissipation}

\author{Rapha\"el Lescanne$^{1,2}$, Samuel Del\'eglise$^3$, Emanuele Albertinale$^{5}$, Ulysse R\'eglade$^{4,1}$, Thibault Capelle$^{3}$, Edouard Ivanov$^{3}$, Thibaut Jacqmin$^3$,  Zaki Leghtas$^{4,1,2}$, Emmanuel Flurin$^{5,3,1}$}
\email{emmanuel.flurin@cea.fr}
\affiliation{$^1$Laboratoire de Physique de l'Ecole Normale Sup\'erieure, ENS, Universit\'e PSL, CNRS, Sorbonne Universit\'e, Universit\'e Paris-Diderot, Sorbonne Paris Cit\'e, Paris, France}
\affiliation{$^2$QUANTIC team, INRIA de Paris, 2 Rue Simone Iff, 75012 Paris, France}
\affiliation{$^3$Laboratoire Kastler Brossel, Sorbonne Universit\'e, CNRS, ENS-PSL Research University,
Coll\`ege de France; 4 place Jussieu, F-75252 Paris, France}
\affiliation{$^4$Centre Automatique et Syst\`emes, Mines-ParisTech, PSL Research University, 60, bd Saint-Michel, 75006 Paris, France}
\affiliation{$^5$Quantronics group, SPEC, CEA, CNRS, Universit\'e Paris-Saclay, CEA Saclay 91191 Gif-sur-Yvette Cedex, France}

\begin{abstract}

Single photon detection is a key resource for sensing at the quantum limit and the enabling technology for measurement based quantum computing. Photon detection at optical frequencies relies on irreversible photo-assisted ionization of various natural materials. However, microwave photons have energies 5 orders of magnitude lower than optical photons, and are therefore ineffective at triggering measurable phenomena at macroscopic scales. Here, we report the observation of a new type of interaction between a single two level system (qubit) and a microwave resonator. These two quantum systems do not interact coherently, instead, they share a common dissipative mechanism to a cold bath: the qubit irreversibly switches to its excited state if and only if a photon enters the
resonator. We have used this highly correlated dissipation mechanism to detect itinerant photons impinging on the resonator. This scheme does not require any prior knowledge of the photon waveform nor its arrival time, and dominant decoherence mechanisms do not trigger spurious detection events (dark counts). We demonstrate a detection efficiency of $58\%$ and a record low dark count rate of $1.4$ per $\mathrm{ms}$. This work establishes engineered non-linear dissipation as a key-enabling resource for a new class of low-noise non-linear microwave detectors. 








\end{abstract}
\date{\today}
\maketitle

\begin{figure}
\includegraphics[width=\columnwidth]{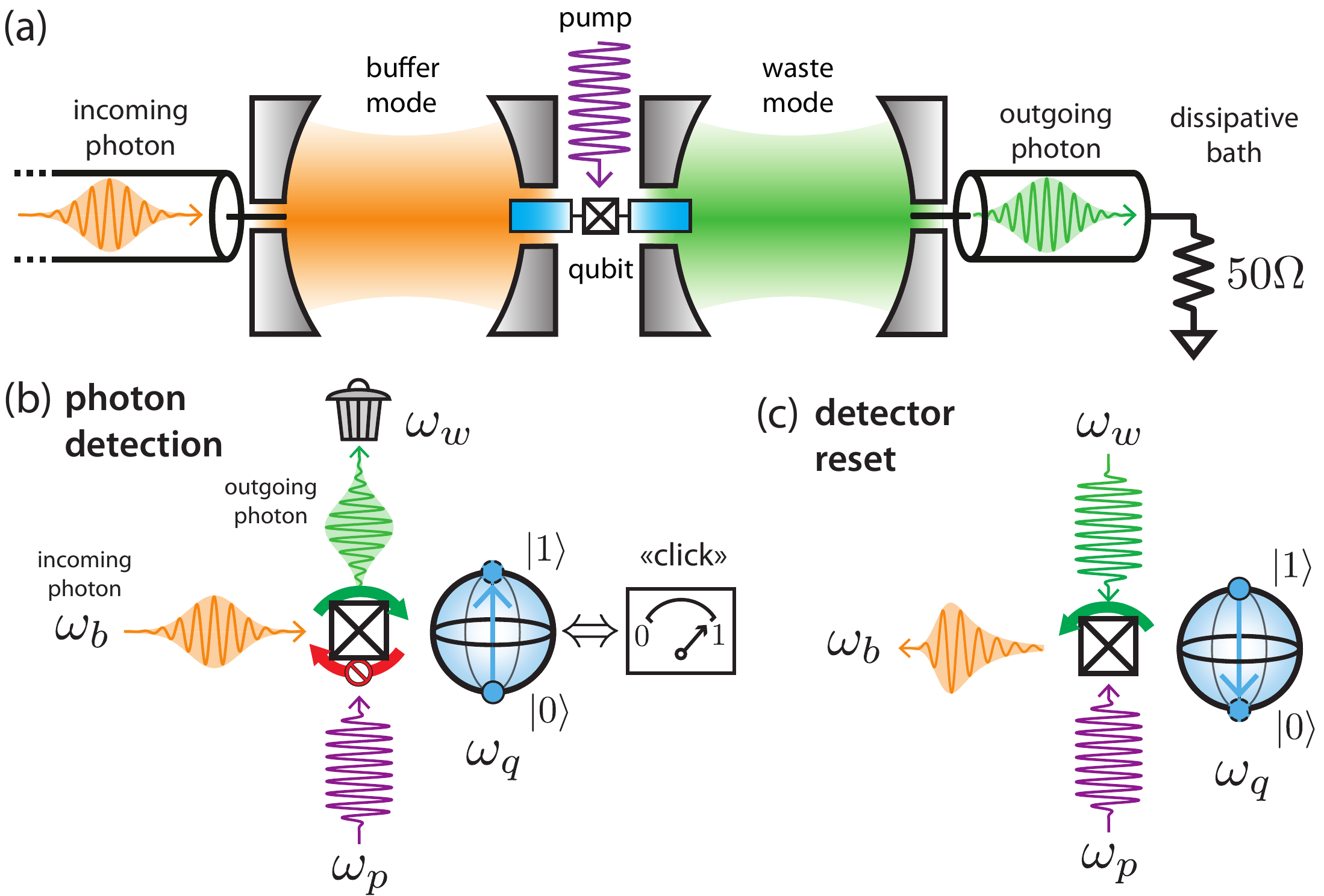}
 \caption{\textbf{Principle of the itinerant photon detector.} (\textbf{a}) The circuit consists of two microwave modes, a buffer (orange) and a waste (green) coupled to a transmon qubit (blue). Each mode is well coupled to its own transmission line so that a photon can enter the device on the left and leave on the right into the dissipative $50\Omega$ environment. A pump (purple) is applied on the Josephson junction to make the three-wave mixing interaction  resonant (Eq. \ref{eq1}). (\textbf{b}) When a photon enters the buffer, the pump converts it via the Josephson non-linearity (black cross) into one excitation of the qubit and one photon in the waste. The waste excitation is irreversibly radiated away in the transmission line, so that the reverse three-wave process cannot occur. The quantum state of the qubit is measured with a standard dispersive readout via the waste to detect whether or not a photon arrived while the pump was on. (\textbf{c}) When a coherent tone drives the waste, the qubit excitation can combine with waste photons and be released via the Josephson non-linearity in the buffer, enabling a fast reset of the detector.}
\label{fig1}
\end{figure}


High-performance photon detectors in the optical domain \cite{Hadfield2009} are the workhorses of various quantum optics experiments: by combining them with Gaussian resources such as optical parametric oscillators or beamsplitters, they have been used to generate non-classical states \cite{Lvovsky2001, Ourjoumtsev2006}, entanglement between remote stationary qubits \cite{Humphreys2018}, entanglement distillation \cite{Kalb2017}, long distance quantum cryptography protocols \cite{Liao2018} or one-way quantum computing \cite{raussendorf2001one}. In such non-deterministic protocols, the preparation of the desired quantum state is heralded by the detection of a single itinerant photon. Crucially, by conditioning the success of the experiment upon the detection events, such protocols are highly robust against losses and inefficiencies whereas their fidelity is directly linked to the number of false-positive events, also known as ``dark-counts'' \cite{Monroe2014}.

Transposing such techniques in the microwave domain\ \cite{helmer2009quantum, romero2009microwave,sathyamoorthy2014quantum, leppakangas2018multiplying, blais2018} would leverage the high level of control over superconducting quantum circuits in a modular architecture where various solid-state quantum systems are connected via lossy connection lines \cite{kimble2008quantum, Chou2018}. Moreover, such detectors are increasingly sought-after due to their applications in the detection of dark matter axions \cite{lamoreaux2013analysis, Brubaker2017}, electron-paramagnetic-resonance spectroscopy \cite{haikka2017proposal}, quantum radars \cite{barzanjeh2015microwave}, or quantum-enhanced imaging \cite{morris2015imaging}. In such applications, where the arrival time of the photon is not known in advance, the possibility to operate the detector in a continuous regime with a high duty-cycle, or equivalently a short reset-time, is a crucial requirement.


In the seemingly unrelated area of dissipation engineering \cite{Kapit2017}, one carefully designs the coupling of a quantum system to a bath to achieve a desired dissipative dynamics. This approach defeats the natural intuition that quantum systems need to be isolated from their environment, and has been used to implement inherently dissipative dynamics, such as the the stabilization of quantum states \cite{Murch2012, Shankar2013, Lin2013} and manifolds \cite{Leghtas2013b, Touzard2018}, and the fabrication of non-reciprocal components that do not rely on an external magnetic field \cite{Metelmann2015, Sliwa2015}.

In this work we use dissipation engineering to couple a single two-level system (qubit) to a transmission line in a peculiar way. An itinerant photon propagating in the line is absorbed by the qubit, but the reverse process is inhibited: an excitation in the qubit does not propagate back in the line. Upon arrival of a single photon, the qubit is left in its excited state, leaving ample time for it to be measured with a microwave pulse containing tens to hundreds of photons, which is measurable with readily available amplification techniques \cite{Macklin2015}. In practice, we demonstrate that implementing this engineered dissipation triggers dark counts at a rate one order of magnitude smaller compared to state-of-the-art experiments \cite{Narla2016, Inomata2016, Opremcak1239, Kono2018, Besse2018}.







The  detector, depicted in Fig.~\ref{fig1}a is composed of two superconducting microwave resonators, the buffer which hosts the incoming field and the waste, which plays the role of the bath, releases the detected photon. The resonators are coupled through a Josephson junction in a bridge transmon configuration \cite{Kirchmair2013}, and are strongly coupled to transmission lines 	at a rate $\kappa_b/2\pi=  1.0 \ \mathrm{MHz}$ and  $\kappa_w/2\pi= 2.4 \ \mathrm{MHz}$, respectively.
A microwave drive, referred to as the pump, is applied to the transmon qubit at frequency 
\begin{equation}
\label{eq:freqmatch}
\omega_p = \bar\omega_q+\omega^e_w-\omega^g_b\;,
\end{equation}
where $\bar\omega_{q}$ is the qubit frequency shifted by the pump power through the  AC-stark effect~\cite{supplement}, $\omega^e_{w}$ and $\omega^g_{b}$  are the buffer and waste frequencies conditioned on the qubit being in its excited state $\ket{e}$ and its ground state $\ket{g}$, respectively. In the absence of the pump $\omega_q/2\pi= 4.532\ \mathrm{GHz}$, $\omega^g_b/2\pi=5.495\ \mathrm{GHz}$ and $\omega^e_w/2\pi=5.770\ \mathrm{GHz}$. The pumped system is well described by the effective Hamiltonian (see \cite{supplement})
\begin{equation}
\hat H_{\mathrm{eff}}/\hbar = g_3 \hat b \hat \sigma^\dagger \hat w^\dagger +  g_3^* \hat b^\dagger \hat \sigma \hat w,
\label{eq1}
\end{equation}
where $g_3$ is the parametrically activated three-wave mixing rate, and verifies $g_3=-\xi_p \sqrt{\chi_{qb} \chi_{qw}}$. Here, $\chi_{qb}/2\pi = 1.02\ \mathrm{MHz}$ and $\chi_{qw}/2\pi = 2.73\ \mathrm{MHz}$ are the dispersive couplings of the buffer and waste to the qubit, respectively. The pump amplitude $\xi_p$ is expressed in units of square root of photons and is typically smaller than one. The buffer and waste annihilation operators are denoted $\hat b$ and $\hat w$, and $\hat \sigma$ denotes the lowering operator of the qubit. The itinerant photon to be detected, incident on the buffer, is converted into a pair of excitations in the qubit and in the waste by the term $\hat b \hat \sigma^\dagger \hat w^\dagger$. We place ourselves in the regime where $|g_3|\ll \kappa_w$ so that the photon in the waste is immediately dissipated in the natural environment while the qubit excitation is stored. Since the waste remains close to its vacuum state throughout the dynamics, the reverse process  ($\hat b^\dagger \hat \sigma \hat w$) is effectively inhibited. The subsequent detection of the qubit in the excited state based on single-shot dispersive readout \cite{Krantz2016} reveals the transit of the photon during the detection time (see Figure \ref{fig1}b).

The irreversible buffer-qubit dynamics, arising from the adiabatic elimination of the waste, is entirely described by a single loss operator \cite{supplement}
\begin{equation}
\hat L=\sqrt{\kappa_\mathrm{nl}}\hat b \hat \sigma^\dagger\;,
\label{eq:lossoperator}\end{equation}
where the engineered dissipation rate is $\kappa_\mathrm{nl}=4 |g_3|^2/\kappa_w$. This dissipator is unusual for two reasons. First, it is non-local \cite{Metelmann2015}, since it involves operators from two different modes. Second, it is non-linear \cite{Leghtas2015}, since it involves the product of these operators. These properties are at the heart of the detection process: under the effect of $\hat L$, the  qubit dissipates towards its excited state conditioned (non-linear) on the buffer (non-local) occupation. As a consequence, the detector is oblivious to the specific mode-shape of the incoming photons (within the detector bandwidth) by evacuating the associated entropy into the unread dissipative channel. 





The efficiency $\eta$ of the detector is defined as the probability of detecting the qubit in its excited state assuming a single incoming photon. For mode-shapes well within the detector bandwidth $(\kappa_\mathrm{nl}+\kappa_b)/2\pi=1.34$ MHz and short compared to the qubit relaxation time, the efficiency is \cite{supplement}
\begin{equation}
\label{eq:eta}
\eta = 4\dfrac{\kappa_\mathrm{nl} \kappa_{b}}{(\kappa_\mathrm{nl}+\kappa_b)^2}.
\end{equation}The efficiency reaches unity for $\kappa_\mathrm{nl}=\kappa_b$, when the non-linear dissipation matches the coupling to the transmission line.

\begin{figure}
\includegraphics[width=\columnwidth]{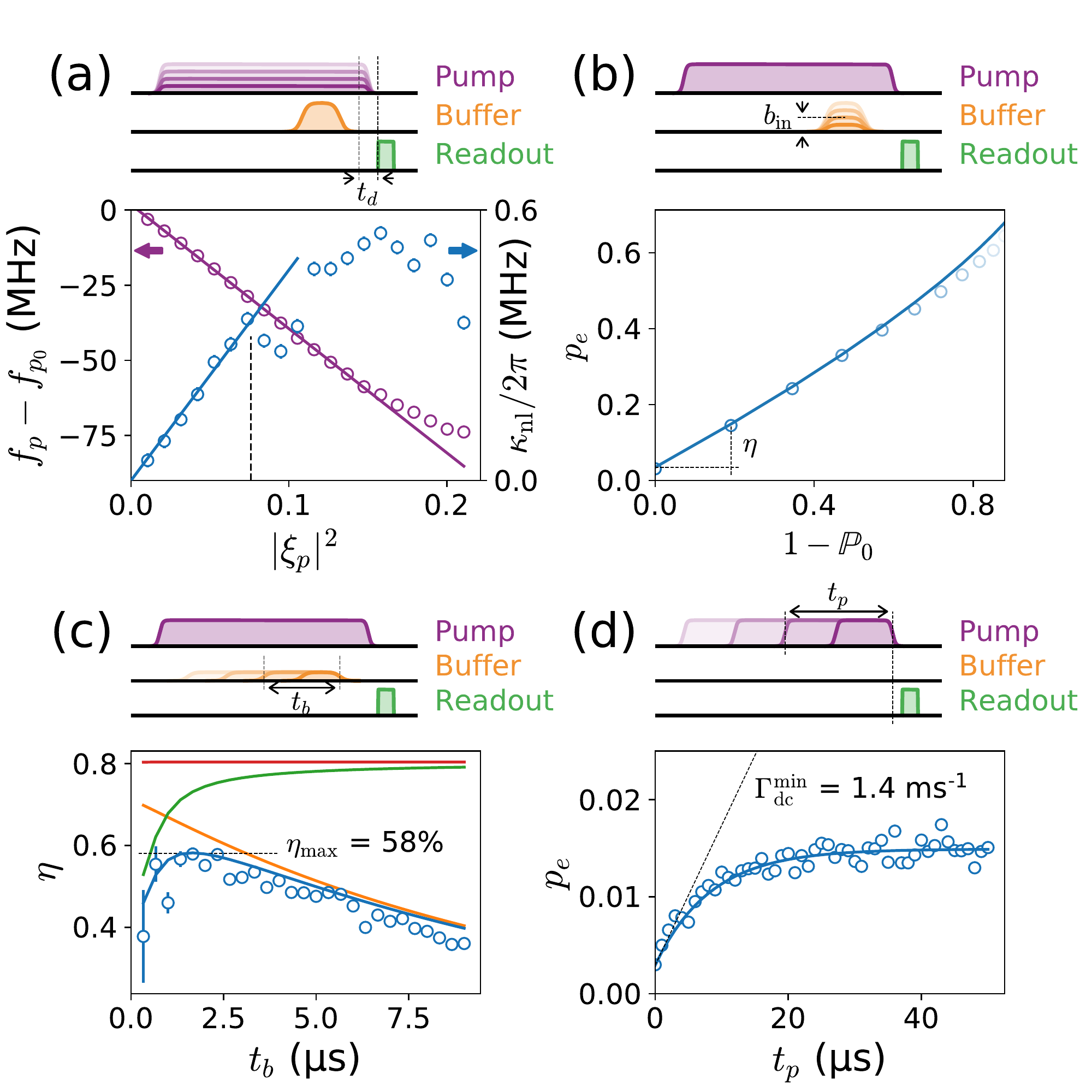}
 \caption{\textbf{Calibration and diagnosis of the detector. }(\textbf{a-d}) A pump tone (purple pulse) of duration $t_p$ powers the device through the qubit mode, and an incoming coherent wave-packet (orange pulse) of duration $t_b$ and amplitude $\bin$ is applied on the buffer. The qubit excited state population $p_e$ is detected after a delay $t_d$ through the waste (green pulse). 
 (\textbf{a}) We plot the pump frequency $f_p$ which maximizes $p_e$ for various pump powers (x-axis). The 10 first data points (purple open circles) are fitted using a linear regression (purple line), leading to a y-intercept within $0.04\%$ of the expected value $f_{p0} = \left(\omega_q+\omega^e_w-\omega^g_b\right)/2 \pi$. The slope of this line provides a calibration of the pump power in units of photon number $|\xi_p|^2$ (see Eq.~\ref{eq:freqmatch2}). For each pump power, we plot the induced non-linear dissipation rate $\kappa_\mathrm{nl}$ (blue open circles), which is extracted from a full model simulation (see Eq.~\eqref{eq:drhoqbdt2}), to best match all measured values of $p_e$ for small values of $b_{in}$ and all values of $t_b$. For pump powers $|\xi_p|^2<0.076$, these extracted values follow the expected linear trend (blue line) $\kappa_\mathrm{nl}=4|\xi_p|^2\sqrt{\chi_{qb}\chi_{wq}}/\kappa_b$, with a $5\%$ deviation at the value fixed for all subsequent experiments $|\xi_p|^2=0.076$ (dashed vertical line). (\textbf{b}) Measured $p_e$ (open circles) as a function of $1-\mathbb{P}_0$, where $\mathbb{P}_0$ is the probability of the incoming wave-packet being in the vacuum. This quantity is varied by increasing $b_{in}$ for a fixed $t_b=2~\mu$s. The slope at the origin is the detection efficiency $\eta$, and the y-intercept is the dark-count. The theory (solid line) is obtained by simulating the full model Eq.~\eqref{eq:drhoqbdt2}. (\textbf{c}) For each $t_b$ (x-axis), $\eta$ is measured (open circles) as in (b). The theory curves (solid lines) are obtained by simulating the full model Eq.~\eqref{eq:drhoqbdt2}, in four regimes: $T_1, \kappa_b=\infty$ (red), $\kappa_b=\infty$ (orange), $T_1=\infty$ (green), and the measured $T_1, \kappa_b$ (blue). Note that the discrepancy between orange and red at small time is due to the finite rising time of the pump pulse $t_d=500\ \mathrm{ns}$. (\textbf{d}) Measured $p_e$ (open circles) as a function of $t_p$ with $\bin=0$, at a repetition rate of $2$~ms. An exponential fit (solid line) rises from $p_e=0.003$ to $p_e = 0.015$ at a rate $1/T_1$.}
\label{fig2}
\end{figure}
In practice, we satisfy Eq.~\eqref{eq:freqmatch} by performing a calibration experiment (Fig. \ref{fig2}a). The pump power is chosen as the largest that did not induce significant qubit heating, thus maximizing the efficiency to dark-count ratio. The chosen pump power leads to $\kappa_\mathrm{nl}/2\pi = 0.370$ MHz which results in an efficiency predicted by Eq.~\eqref{eq:eta} of $80\%$ (red line of Fig.~\ref{fig2}c). The detector efficiency is measured by varying the amplitude of a calibrated coherent pulse \cite{LescannePRApp2019} as shown in Fig.~\ref{fig2}b. The measurement is repeated for increasing pulse length as shown in Fig.~\ref{fig2}c. We observe a smooth dependence of the efficiency as a function of the pulse length, that is well reproduced by our model if we take into account the finite detector bandwidth at short pulse duration (green curve of Fig.~\ref{fig2}c), and the finite qubit lifetime for long pulses (orange curve of Fig.~\ref{fig2}c), resulting in a maximum detection efficiency of $\eta_\mathrm{max}=58\%$  for a 2 $\mu$s pulse length.



\begin{figure}
\includegraphics[width=\columnwidth]{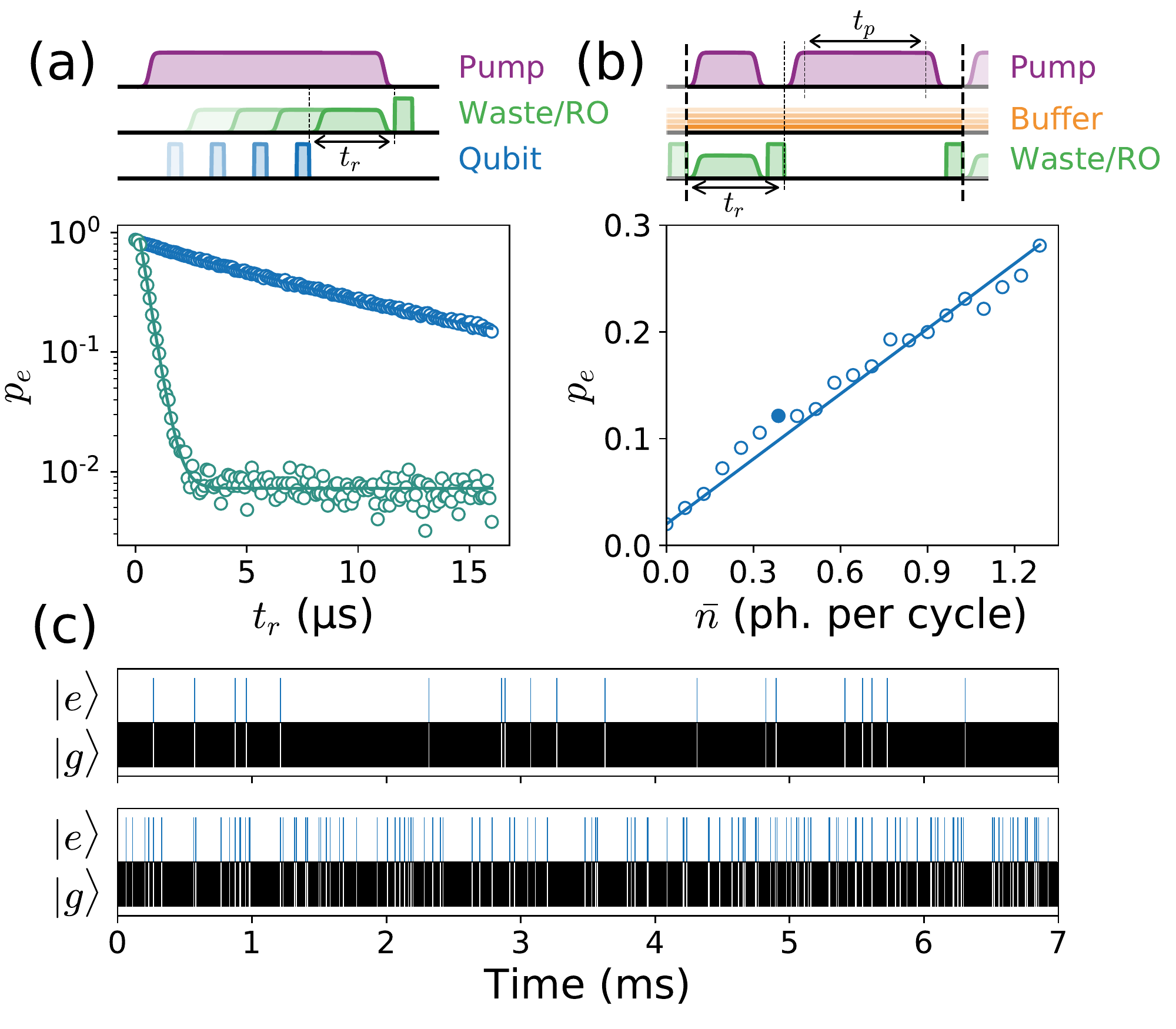}
 \caption{\textbf{Reset protocol and cyclic operation. }(\textbf{a}) The qubit is initialized in its excited state (blue pulse) and the pump is activated. The decay of $p_e$ is measured (open circles) in the presence (green) or absence (blue) of a reset tone of duration $t_r$ applied to the waste mode (green pulse). The exponential fits (solid lines) yield a qubit $T_1$ reduced from $7.7~\mu$s to $370~$ns. Note that the residual qubit population is reduced in the presence of the reset tone from $3.5 \%$ (see Fig.~\ref{fig2}) to $0.7 \%$, demonstrating that the reset is in fact cooling the qubit below its effective bath occupation. (\textbf{b}) Cyclic operation of the pulse sequence in between the two vertical dashed lines, in the presence of a continuous input tone of variable amplitude (orange pulse). First, the qubit is reset for $2~\mu$s (first green and purple pulses), followed by a heralding measurement to confirm the preparation in the ground state (second green pulse). The detector is then activated during a time $t_p=3~\mu$s by switching on the pump (second purple pulse), and finally, the qubit state is readout (third green pulse). The measured $p_e$ (open circles) is plotted as a function of the photon number in the input pulse integrated over the cycle time $7~\mu$s. A linear regression (solid line) yields an overall efficiency of $20~\%$ (which accounts for the $43\%$ duty cycle), and a dark count rate of $2~\%$. (\textbf{c}) Real-time trajectories of 1000 successive detection cycles, with $\bar{n}=0$ (top) and $\bar{n}=0.38$ (bottom), as marked by the full dot in (b).}
\label{fig3}
\end{figure}

A crucial figure of merit of the detector is the dark count rate $\Gamma_\mathrm{dc}$, defined as the number of clicks per unit-time in the absence of incoming photons. 
By virtue of our dissipation engineering approach, $\Gamma_\mathrm{dc}$ is robust against the two main decoherence mechanisms of the qubit: dephasing and energy relaxation. This is in stark contrast with the most advanced schemes relying on Ramsey interferometry \cite{Kono2018,Besse2018}, where the dark count rate scales with the qubit dephasing rate.
Fig.~\ref{fig2}d shows the detection probability $p_e$ as a function of the detection window $t_p$ in the absence of buffer excitation. At short time compared to the qubit relaxation time $T_1$, we find $p_e=0.003+\Gamma_\mathrm{dc}\times t_p$. The first term of 0.003 results from the improper initialization of the qubit in its ground state and detection errors. We find $\Gamma_\mathrm{dc} = 1.4  \,\mathrm{ms}^{-1}$, limited by pump-induced heating of the qubit bath \cite{supplement, LescannePRApp2019, VerneyPRApp2019} (see Fig.\ref{fig2}d).  
We have observed that for the same detection window, when the repetition rate increases from $(2\  \mathrm{ms})^{-1}$ to $(50\ \mathrm{\mu s})^{-1}$ , the dark-count rate increases to $4~\mathrm{ms}^{-1}$, suggesting that the qubit bath thermalizes over timescales of hundreds of microseconds.
In spite of the modest lifetime of our transmon qubit, the dark count rate reported here is one order of magnitude lower than the values reported previously in the literature \cite{Narla2016, Inomata2016, Kono2018, Besse2018}.
A key requirement for practical photon detectors is a fast and high-fidelity reset, which we achieve using the reverse process $g_3^* \hat b^\dagger \hat \sigma \hat w$ (see Fig.~\ref{fig1}c). By shining a resonant tone on the waste in the presence of the pump, the qubit relaxes to the ground state with the emission of a photon in the buffer, leading to a fast reset at a rate  $\gamma_\mathrm{reset}=(370\ \mathrm{ns})^{-1}$ and a ground state preparation fidelity of $F^\mathrm{reset}_{|g\rangle}=99.3\%$ improved to $F^\mathrm{herald}_{|g\rangle}=99.6\%$ with a subsequent heralding readout (Fig.~\ref{fig3}a). The down time of the detector comprises the reset, heralding and readout pulses, and is as low as $3.5\ \mathrm{\mu s}$ (Fig.~\ref{fig3}b). The continuous operation is demonstrated, by running the detector with a $43\%$ duty cycle in the presence of a coherent tone on the buffer of variable strength (Fig.~\ref{fig3}c). Moreover, the qubit is at all times either in $\ket{g}$ or $\ket{e}$ \cite{supplement}, and therefore a weak and continuous dispersive readout could be added to detect uninterruptedly the incoming photons with a precise timing of the clicks.

\begin{figure}
\includegraphics[width=\columnwidth]{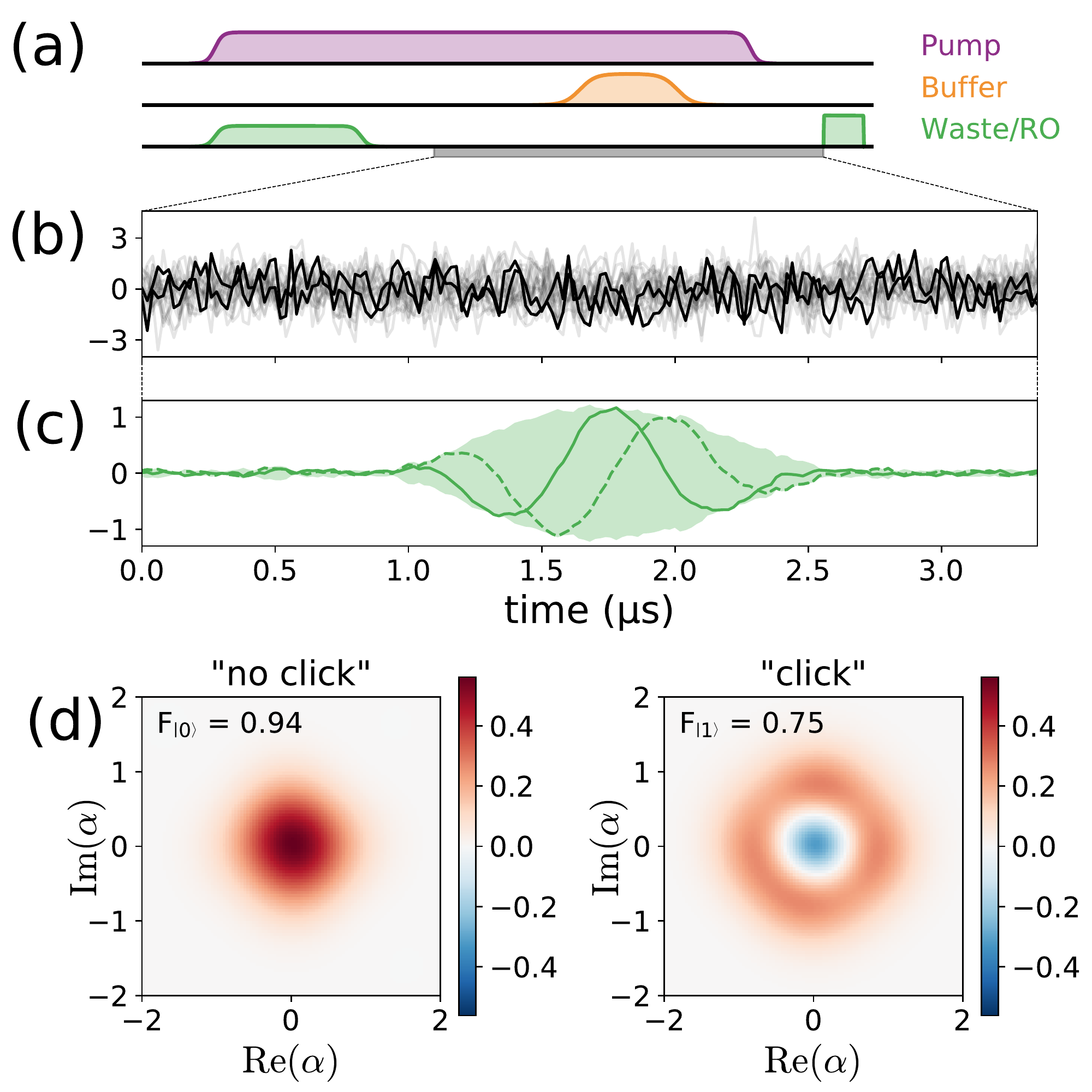}
 \caption{\textbf{Quantum non-demolition characterization by tomography of the quantum state released in the waste line.
 }  (\textbf{a}) The three wave-mixing interaction is activated by a pump tone (purple pulse) during which the detector is first initialized by a reset pulse on the waste port (first green pulse) preceding the injection of a small coherent field (mean occupation $\bar n = 0.35$) on the buffer port. The qubit excitation, detected at the end of the sequence by a readout pulse (second green pulse) is correlated with the creation of a single photon by the waste. During the detection process, a near quantum-limited phase-preserving amplifier is used to monitor the field emitted by the waste (grey-shaded rectangle). (\textbf{b}) The signal, demodulated at the frequency $(\omega_w^g+\omega_w^e)/2$, is dominated by quantum and amplifier noise, as exemplified with 20 overlayed single-quadrature time-traces (2 time traces have been coloured in black for readability). (\textbf{c}) The waveform of the emitted wavepacket can be deconvoluted from the uncorrelated detection noise by statistical analysis of the traces \cite{Morin2013}. The real, imaginary part, and envelope of the waveform retrieved experimentally are represented in full line, dashed line, and green shaded area respectively. (\textbf{d}) The quantum state of the field in the temporal mode plotted in (c) is reconstructed by analyzing the moments of the measured amplitude distribution \cite{Eichler2011}. The Wigner function of the state conditioned on the detection of the qubit in state $|g\rangle$ (resp. $|e\rangle$) is represented on the left (resp. right).}
\label{fig4}
\end{figure}

Finally, the waste which up to now has been considered as a bath, can in fact be monitored to demonstrate that our detector is quantum non-demolition. Indeed, the qubit acts as a witness of the photon passage, the latter is first absorbed in the buffer and released through the waste at another carrier frequency (Fig.~\ref{fig1}b). By performing the heterodyne detection of the outgoing field, heralded by the click of the detector, we first extract the temporal mode of the photon based on the eigen-mode expansion of the auto-correlation function \cite{Morin2013}. Then, we reconstruct the density matrix of the photon by measuring the moments of the signal distribution up to the fourth order \cite{Eichler2011}.
Based on a maximum likehood reconstruction, we obtain a single photon fidelity $F_{|1\rangle} = 75\ \%$ conditioned on a click of the detector and a vacuum fidelity $F_{|0\rangle}=94\%$ conditioned on the absence of a click (Fig. \ref{fig4}).

We have realized a non-linear and non-local dissipator involving a qubit coupled to a resonator. Conditioned on the impact of a photon on the resonator, the qubit irreversibly switches to its excited state which is then detected with a single-shot readout. This photon detector achieves both high efficiency and an unprecedentedly low dark count rate owing to its robustness against the main decoherence mechanisms of superconducting circuits. Alike Josephson parametric amplifiers, this new class of robust detectors based on dissipation engineering constitute an essential step towards the practical use of photon detectors in the microwave domain. 
By engineering higher-order dissipators, one can envision more complex detection patterns, such as number-resolving or multi-mode correlation detectors.




\paragraph{Acknowledgements}
ZL acknowledges support from the ANR grant ENDURANCE, and the EMERGENCES grant ENDURANCE of Ville de Paris. SD acknowledges support from the ANR grant QuNaT. EI acknowledges support from the European Union's Horizon 2020
Programme for Research and Innovation under grant agreement No. 722923
(Marie Curie ETN - OMT). The devices were fabricated within the consortium Salle Blanche Paris Centre. EF aknowledges support the ENS Junior Research Chair under grant LabEX ENS-ICFP: ANR-10-LABX-0010/ ANR-10-IDEX-0001-02 PSL*.

\bibliography{biblio}
\bibliographystyle{ieeetr}

\newpage

\clearpage

\onecolumngrid

\section{Supplementary materials}

\subsection{Circuit parameters}
\begin{figure}[h!]
\includegraphics[width={100mm}]{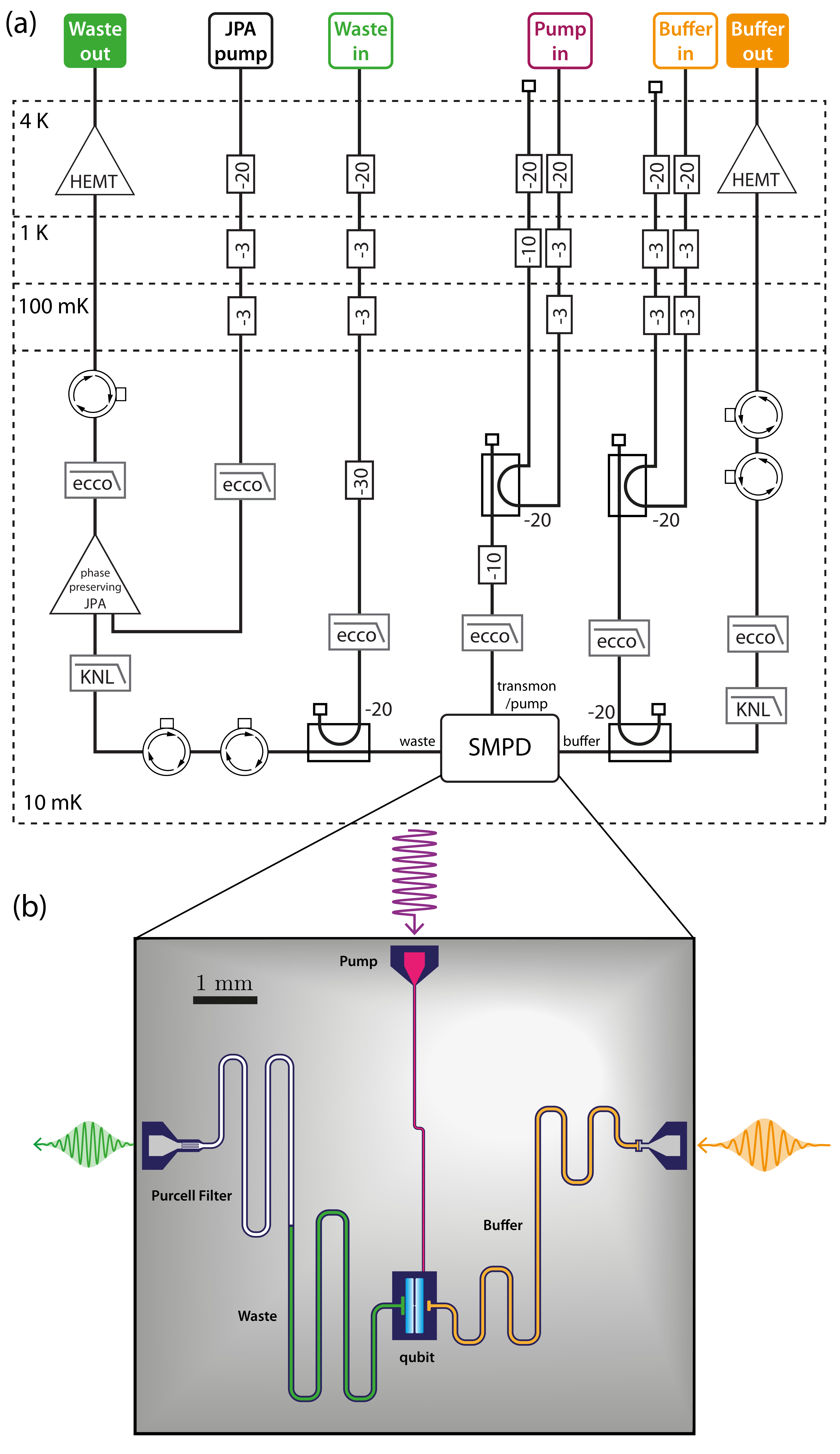}
 \caption{(a) Schematic of microwave cabling of the dilution refrigerator. (b) Superconducting coplanar-waveguide layout, note that this vector layout is the exact mask used for the circuit fabrication.}
\label{fig_circuit}
\end{figure}

The circuit consists of $\lambda/2$-coplanar waveguide resonators as depicted in Fig.\ref{fig_circuit}(b). The circuit is made out of sputtered Niobium with a thickness of  $120 \,\mathrm{nm}$ deposited on a $280 \,\mathrm{\mu m}$-thick wafer of intrinsic silicon.  The main circuit is etched after an optical lithography step and then the Josephson junction is made of evaporated aluminum through a PMMA/MAA resist mask written in a distinct ebeam lithography step.

\begin{center}
\begin{table}[h!]
    \begin{tabular}{| l | l | l  |}
    \multicolumn{2}{c}{qubit}       \\ \hline
    $\omega_q/2\pi$& $4.532\ \mathrm{GHz}$ \\ \hline
    $T_1$ & $8-9\ \mathrm{\mu s}$  \\ \hline
    $T_{2}^*$ & $10\ \mathrm{\mu s}$   \\ \hline
    $\chi_{qq}/2\pi$& $146\ \mathrm{MHz}$ \\ \hline
    $\chi_{qb}/2\pi$& $1.02\ \mathrm{MHz}$ \\ \hline
    $\chi_{qw}/2\pi$& $2.73\ \mathrm{MHz}$ \\ \hline
    \multicolumn{2}{c}{ Waste mode }       \\ \hline
    $\omega^g_b/2\pi$& $5.4952\ \mathrm{GHz}$ \\ \hline
    $\kappa_\mathrm{ext}/2\pi$& $ 2.38\ \mathrm{MHz}$  \\ \hline
    $\kappa_\mathrm{int}/2\pi$& $ < 100\ \mathrm{kHz}$  \\ \hline
      \multicolumn{2}{c}{ Buffer mode }         \\ \hline
    $\omega^g_w/2\pi$& $5.7725\ \mathrm{GHz}$ \\ \hline
    $\kappa_\mathrm{ext}/2\pi$ & $0.890\ \mathrm{MHz}$  \\ \hline
    $\kappa_\mathrm{int}/2\pi$ & $80\ \mathrm{kHz}$  \\ \hline
    
    \end{tabular}
    \caption{Measured system parameters. Note that the internal losses of the waste are too small in comparison to the external losses to be accurately measured.}
    \label{table}
    \end{table}
    
\end{center}

\subsection{Purcell Filter}

In order to strongly couple the waste resonator to the cold $50\ \Omega$ environment which provides the dissipative character of the scheme without degrading the qubit relaxation time $T_1$, we have employed a Purcell filter in series with the waste resonator as depicted in Fig.\ref{fig_circuit}(b). As described in ref.~\cite{sete2015quantum}, another benefit of the Purcell filter is that the  readout time of the qubit is reduced hence the readout fidelity is enhanced.

The Purcell filter is a bandpass filter consisting in a $\lambda/2$ resonator coupled in series with the waste resonator at a rate $\mathcal{G}$ and placed on resonance with the latter, $\omega_\mathrm{Purcell}=2\pi\times5.786\ \mathrm{GHz}$, and coupled to the transmission line at a rate $\kappa_\mathrm{Purcell}=2\pi\times36\ \mathrm{MHz}$. The filter is designed to be in the strong damping regime $\kappa_\mathrm{Purcell}\gg\mathcal{G}$ so that the expected anti-crossing due to the hybridization of the modes is irrelevant.
 The effective decay rate of the waste resonator to the transmission line through the filter is then given by:
\begin{equation}
\kappa_w=\dfrac{4\mathcal{G}^2}{\kappa_\mathrm{Purcell}}\dfrac{1}{1+[2(\omega_\mathrm{Purcell}-\omega_w)/\kappa_\mathrm{Purcell}]^2}\approx\dfrac{4\mathcal{G}^2}{\kappa_\mathrm{Purcell}}
\end{equation}
Experimentally, we extract an effective decay rate of $\kappa_w=2\pi\times2.38\ \mathrm{MHz}$ leading to a coupling strength $\mathcal{G}=2\pi\times5.6\ \mathrm{MHz}$ between the waste and the Purcell filter.
The qubit is coupled to the waste resonator at a rate $g=2\pi\times41\ \mathrm{MHz}$. Its residual decay rate through the waste channel in presence of the Purcell filter is given by:
\begin{equation}
\kappa_q^{(w)}=\dfrac{g^2}{(\omega_w-\omega_q)^2}\dfrac{4\mathcal{G}^2}{\kappa_\mathrm{Purcell}}\dfrac{1}{1+[2(\omega_\mathrm{Purcell}-\omega_q)/\kappa_\mathrm{Purcell}]^2}\approx\dfrac{g^2\mathcal{G}^2}{(\omega_w-\omega_q)^4}\kappa_\mathrm{Purcell}=2\pi\times1\ \mathrm{Hz}.
\end{equation}
Note that for a given decay rate of the waste in the transmission line, the Purcell filter enables the suppression of the qubit relaxation thought the waste channel by a factor $\kappa_\mathrm{Purcell}^2/(\omega_\mathrm{Purcell}-\omega_q)^2\sim  10^{-3}$.

\clearpage

\subsection{System Hamiltonian derivation}

Our system consists of three electromagnetic modes, referred to as the buffer, the waste, and the qubit, coupled through a Josephson junction. A strong radio-frequency drive, referred to as the pump, is applied to the qubit mode. The Hamiltonian of this system is well described by
\begin{eqnarray*}
\hat{H}/\hbar & = & \sum_{m=b,w,q}\omega_m \hat{m}^\dag\hat{m}-\frac{E_J}{\hbar}\left(\cos\left(\hat\varphi\right)+\hat{\varphi}^2/2\right)+2\epsilon_p\cos(\omega_pt)(\hat{q}+\hat{q}^\dag)\;,\\
\hat{\varphi} & = & \sum_{m=b,w,q}\varphi_m(\hat{m}+\hat{m}^\dag)
\end{eqnarray*}
where the index $m=b,w,q$ refers to the buffer, waste and qubit modes respectively, of angular frequency $\omega_m$, and annihilation operator $\hat{m}$. The Josephson energy is denoted $E_J$, and  $\hat\varphi$ is the phase across the junction, which can be decomposed as the sum of the phase $\varphi_m (\hat{m}+\hat{m}^\dag)$ across each mode, where $\varphi_m$ is the zero point fluctuation of the phase across mode $m$. The pump's amplitude and angular frequency are denoted $\epsilon_p$ and $\omega_p$ respectively.

The strong pump is accounted for by moving into a frame where the qubit mode is displaced by its mean field amplitude $\xi_p e^{-i\omega_p t}$ where $\xi_p\approx-\epsilon_p /(\omega_q-\omega_p)$ \cite{Leghtas2015}, provided that the pump is adiabatically switched on and off with respect to the detuning $\omega_q-\omega_p$ (about 300 MHz in our experiment) \cite{Touzard2018}. 

Following this frame displacement, we place ourselves in the interaction picture with respect to the Hamiltonian $\sum_{m=b,w,q}(\omega_m-\delta_m) \hat{m}^\dag\hat{m}$, where $\delta_m$ are arbitrary detunings that will be used to cancel the AC Stark-shifts due to the Kerr effect. The transformed Hamiltonian reads:

\begin{eqnarray*}
\hat{H'}/\hbar & = &\sum_{m=b,w,q}\delta_m\hat{m}^\dag\hat{m} -\frac{E_J}{\hbar}\left(\cos\left(\hat\varphi'\right)+\hat{\varphi'}^2/2\right)\;,\\
\hat{\varphi'} & = & \sum_{m=b,w,q}\varphi_m\left(\hat{m}e^{-i(\omega_m-\delta_m)t}+\hat{m}^\dag e^{i(\omega_m-\delta_m)t}\right) + \varphi_q\left(\xi_p e^{-i\omega_p t}+\xi_p^* e^{i\omega_p t}\right)\;.
\end{eqnarray*}

We now take
\begin{equation}
\label{eq:freqmatch_supp_mat}
\omega_p = (\omega_q-\delta_q)+(\omega_w-\delta_w)-(\omega_b-\delta_b)\;,
\end{equation}
and expanding the cosine term to 4th order, and keeping only the non-rotating terms leads to

\begin{eqnarray}
\label{eq:H'}
\hat{H}'&\approx& H_{\text{Stark}} + H_\text{Kerr}+H_{4 WM}\\
\hat{H}_{\text{Stark}}/\hbar &=&\sum_{m=b,w}{(\delta_m-\chi_{qm}|\xi_p|^2)\hat{m}^\dag\hat{m}} + (\delta_q-2\chi_{qq}|\xi_p|^2)\hat{q}^\dag\hat{q}\\
\hat{H}_\text{Kerr}/\hbar &=& \sum_{m=b,w, q}-\frac{\chi_{mm}}{2}{\hat{m}^\dag}{}^2{\hat{m}}{}^2 - \chi_{qb}\hat{b}^\dag\hat{b}\hat{q}^\dag\hat{q} - \chi_{qw}\hat{w}^\dag\hat{w}\hat{q}^\dag\hat{q} - \chi_{bw}\hat{b}^\dag\hat{b}\hat{w}^\dag\hat{w}\\
\hat{H}_{4WM}/\hbar &=& g_3 \hat{b}\hat{w}^\dag\hat{q}^\dag + g_3^* \hat{b}^\dag\hat{w}\hat{q}
\end{eqnarray}
With the coefficients $\hbar \chi_{mm} = E_J  \phi_m^4/2$, $\hbar \chi_{qb} = E_J \phi_q^2 \phi_b^2$, $\hbar \chi_{qw} = E_J \phi_q^2 \phi_w^2$, $\hbar \chi_{bw} = E_J \phi_b^2 \phi_w^2$ and $\hbar g_3 = -E_J \xi_p \phi_q^2 \phi_b \phi_w$, which can also be written as:
\begin{equation}
g_3 = -\xi_p \sqrt{\chi_{qb} \chi_{qw}}\;.
\end{equation}

Note that we have neglected terms of the form $\hat{m}^\dag\hat{m}$ arising from the normal ordering of the 4$^\mathrm{th}$ order term, since they simply shift the bare frequencies $\omega_m$ by a constant amount. Since the qubit anharmonicity $\chi_{qq}$ is much larger than all the dissipation and excitation rates, in the following, we project the qubit mode onto its two lowest energy levels $\ket{g}$ and $\ket{e}$. We thus replace the bosonic operator $\hat{q}$ by the two-level lowering operator $\hat\sigma = \ket{g}\bra{e}$. Moreover, we choose the mode reference frames such that $\delta_q = 2\chi_{qq}|\xi_p|^2, \delta_b = \chi_{qb}|\xi_p|^2$ and $\delta_w=\chi_{qw}|\xi_p|^2+\Delta$. We have introduced an arbitrary detuning $\Delta$ which, as we will see in the next section, can be chosen to cancel the effect of the cross-Kerr between the qubit and the waste $\chi_{qw}$. This leads to $\hat H_\mathrm{Stark}/\hbar=\Delta \hat{w}^\dag \hat{w}$. The pump frequency thus needs to be adapted for each value of $\xi_p$ in order to always verify \eqref{eq:freqmatch_supp_mat}:
\begin{equation}
\label{eq:freqmatch2}
\omega_p = \omega_q + \omega_w -\omega_b - \Delta -|\xi_p|^2\left(2\chi_{qq}+\chi_{qw}-\chi_{qb}\right)\;.
\end{equation}
The Hamiltonian now reads:
\begin{align}
    \hat{H}''/\hbar &=g_3 \hat{b}\hat{w}^\dag\hat{\sigma}^\dag +g_3^* \hat{b}^\dag\hat{w}\hat{\sigma}+\Delta\hat{w}^\dag\hat{w} \nonumber\\
&+  \sum_{m=b,w}-\frac{\chi_{mm}}{2}{\hat{m}^\dag}{}^2{\hat{m}}{}^2 - \chi_{qb}\hat{b}^\dag\hat{b}\hat{\sigma}^\dag\hat{\sigma} - \chi_{qw}\hat{w}^\dag\hat{w}\hat{\sigma}^\dag\hat{\sigma} - \chi_{bw}\hat{b}^\dag\hat{b}\hat{w}^\dag\hat{w}
\label{Hamiltonian_full}
\end{align}

\subsection{Adiabatic elimination of the waste mode}
A crucial resource in our system is the intentional dissipation of the waste mode $w$, which is coupled to a transmission line with a rate $\kappa_w$. In addition, the buffer mode is over-coupled (rate $\kappa_w$) to the input line which carries the incoming photons. Finally, the qubit mode is designed to be isolated from its environment, but inevitably has a residual uncontrolled dissipation at a rate $\kappa_q$, and dephasing $\kappa_\phi$.
\begin{eqnarray}
\frac{d}{dt} \rho&=& -i[\hat{H}''/\hbar,\rho]+\kappa_w\DD[\hat{w}]\rho+\kappa_b\DD[\hat{b}]\rho+\kappa_q\DD[\hat{\sigma}]\rho+\frac{\kappa_\phi}{2}\DD[\hat{\sigma}_z]\rho
\label{full_dynamics}
\end{eqnarray}
where the Lindblad operator is defined for any operator $\hat{O}$ as $\DD[\hat{O}]\rho = \hat{O}\rho\hat{O}^\dag-\frac{1}{2}\hat{O}^\dag\hat{O}\rho-\frac{1}{2}\rho\hat{O}^\dag\hat{O}$

We place ourselves in the regime where $|g_3|, \chi_{qb}, \chi_{bw}, \chi_{bb}, \kappa_b, \kappa_q,\kappa_\phi \sim \delta\kappa_w$, and $\delta$ is a small parameter $\delta \ll 1$. In our experiment, $\chi_{qw}/\kappa_w \sim 1$, and we assume $\Delta/\kappa_w\sim 1$. In this regime, the waste mode can be adiabatically eliminated, leading to an effective dynamics for the buffer and qubit modes alone. Following \cite{Leghtas2015}, we search for a solution of the form
\begin{equation*}
\rho = \rho_{00}\ket{0}\bra{0}+\delta\left(\rho_{10}\ket{1}\bra{0}+\rho_{01}\ket{0}\bra{1}\right)+\delta^2\left(\rho_{11}\ket{1}\bra{1}+\rho_{20}\ket{2}\bra{0}+\rho_{02}\ket{0}\bra{2}\right)+O(\delta^3)\;,
\end{equation*}
and we are interested in the dynamics of the reduced density operator for the qubit-buffer modes, which is obtained by taking the partial trace over the waste mode: $\rho_{qb} = \text{Tr}_{w}(\rho) = \rho_{00}+\delta^2\rho_{11}$. We rewrite the Hamiltonian of Eq. \eqref{Hamiltonian_full} in the following form
\begin{eqnarray}
\label{eq:Hpp}
    \hat{H}''/\hbar &=& g_3 \hat{b} \hat{\sigma}^\dag\hat{w}^\dag + g_3^* \hat{b}^\dag \hat{\sigma}\hat{w} + \left(\Delta - \chi_{qw}\hat{\sigma}^\dag\hat{\sigma}- \chi_{bw}\hat{b}^\dag\hat{b}\right)\hat{w}^\dag\hat{w} + \hat{H}_{bq}/\hbar\\
    \hat{H}_{qb}/\hbar &=& -\frac{\chi_{bb}}{2}{\hat{b}^\dag}{}^2{\hat{b}}{}^2 - \chi_{qb}\hat{b}^\dag\hat{b}\hat{\sigma}^\dag\hat{\sigma}
\end{eqnarray}
and we define
\begin{equation}
\mathcal{L}_{qb}(\rho_{qb}) = -\frac{i}{\hbar}[\hat{H}_{qb},\rho_{qb}] + \kappa_b\DD[\hat{b}]\rho_{qb}+\kappa_q\DD[\hat{\sigma}]\rho_{qb}+\frac{\kappa_\phi}{2}\DD[\hat{\sigma}_z]\rho_{qb}
\end{equation}
By multiplying Eq. \eqref{full_dynamics} by $\bra 0 ...\ket0$, $\bra0 ... \ket1$, and $\bra1...\ket1$ respectively, we get:

{
\begin{eqnarray}
\frac{d}{\kappa_w dt}\rho_{00}&=& \delta^2 \left( i\rho_{01}\hat{A}-i\hat{A}^\dag\rho_{10}+ \rho_{11} \right) +\frac{1}{\kappa_w}\mathcal{L}_{qb}(\rho_{00})+ O(\delta^3)\label{eq:drho00} \\
\frac{d}{\kappa_w dt}\rho_{01}&=&  i\rho_{00}\hat{A}^\dag-\rho_{01} \left(\frac{1}{2} - i\hat{\Delta} \right)+ O(\delta) \label{eq:drho01}\\
\frac{d}{\kappa_w dt}\rho_{11}&=&  i\rho_{10}\hat{A}^\dag- i\hat{A}\rho_{01} -i [\hat{\Delta},\rho_{11}] - \rho_{11} + O(\delta)\label{eq:drho11}
\end{eqnarray}
}
where
\begin{eqnarray}
\hat{A} &=& \frac{g_3} {\kappa_w \delta}\hat{b} \hat{\sigma}^\dag\\
\hat{\Delta} &=& \frac{\Delta-\chi_{qw}\hat{\sigma}^\dag\hat{\sigma}}{\kappa_w}
\end{eqnarray}
Note that $||\hat{A}||$ and $||\hat\Delta||$ are of order $\delta^0$. Considering Eq.~\eqref{eq:drho01}, we see that the derivative of $\rho_{01}$ is composed of a term proportional to $\rho_{00}$ that can be viewed as an external drive, and a term proportional to $\rho_{01}$, that includes a damping term. Since the variation of $\rho_{00}$ is slow ($d\rho_{00}/\kappa_w dt$ of order $\delta^2$, see Eq.~\eqref{eq:drho00}) in comparison to the damping term (of order 1), we can make the adiabatic approximation: we consider that $\rho_{01}$ is continuously in its steady state. The same reasoning applies to $\rho_{11}$, we thus set to 0 the left hand sides of Eqs. \ref{eq:drho01} and \ref{eq:drho11}. Moreover, by noting that $\hat\Delta\hat A = (\Delta-\chi_{qw})\hat{A}/\kappa_w$, we can solve for $\rho_{01}, \rho_{10}, \rho_{11}$ as a function of $\rho_{00}$. We find

\begin{eqnarray}
\label{eq:rho01}
    \rho_{01} &=& \frac{1}{1+4|\frac{\Delta-\chi_{qw}}{\kappa_w}|^2}\left(2i-4\frac{(\Delta-\chi_{qw})}{\kappa_w}\right)\rho_{00}\hat{A}^\dag\\
    \label{eq:rho11}
    \rho_{11} &=& \frac{1}{1+4|\frac{\Delta-\chi_{qw}}{\kappa_w}|^2} 4\hat{A}\rho_{00}\hat{A}^\dag \;.
\end{eqnarray}

We denote

\begin{eqnarray}
\kappa_\mathrm{nl} &=& \frac{4|g_3|^2/\kappa_w}{1+4|\frac{\Delta-\chi_{qw}}{\kappa_w}|^2} \; ,\\
\Delta_{nl} &=& \frac{4|g_3|^2/\kappa_w}{1+4|\frac{\Delta-\chi_{qw}}{\kappa_w}|^2}\frac{\chi_{qw} - \Delta}{\kappa_w} \;.
\end{eqnarray}

Inserting the solutions \eqref{eq:rho01}, \eqref{eq:rho11} into Eq.~\eqref{eq:drho00} we find

\begin{eqnarray}
\label{eq:drhoqbdt}
    \frac{d}{dt}\rho_{00} &=& -i\Delta_{nl}[\hat b^\dag\hat b \hat \sigma \sigma^\dag, \rho_{00}] + \kappa_\mathrm{nl}\mathcal{D}[\hat b\hat\sigma^\dag]\rho_{00} + \mathcal{L}_{qb}(\rho_{00}) + O(\delta^3)\;.
\end{eqnarray}

The term proportionnal to $\kappa_\mathrm{nl}$ is the non-linear damping term at the heart of the Single Microwave Photon Detector. It is maximized for $\Delta = \chi_{qw}$. In this configuration, the pump angular frequency $\omega_p = \omega_q + \omega_w - \omega_b - \chi_{qw}$ is such that $\hbar \omega_p$ exactly matches the energy difference between the initial state $\hat{b}^\dag \ket 0$ and final state $\hat{\sigma}^\dag\hat{w}^\dag\ket0$, and $\kappa_\mathrm{nl} = 4 |g_3|^2/\kappa_w$. The term proportional to $\Delta_{nl}$ is a ``generalized frequency pull'' that corresponds to a tunable cross-Kerr effect between the qubit and buffer modes. Note that the $\rho_{qb}$ follows the same dynamics as Eq.~\eqref{eq:drhoqbdt}.

\subsection{Qubit dynamics and detection efficiency}
\subsubsection{Efficiency for single-photon Fock states}
The detector efficiency is defined as the probability $p_e$ to find the qubit in the excited state, when a Fock state $\ket{1}$ is incident on the buffer cavity. However in this work, we calibrate our photon-detector with coherent states. In the following, we provide a simple argument to bridge the gap between this definition of $\eta$ and the experiment performed with coherent states.

We can formally describe the output of a random source that emits a single-photon with a small probability $\epsilon$ with the density matrix 
\begin{equation}
    \rho = (1-\varepsilon)\ket{0}\bra{0}+\varepsilon\ket{1}\bra{1}\; .
\end{equation}
With such a source, the detector should click with a probability $p_e =\eta\epsilon$.
Besides, a coherent state with a small complex amplitude $\alpha$ reads:
\begin{equation}
    |\psi\rangle \approx (1-\frac{1}{2}|\alpha|^2)(\ket{0}+\alpha\ket{1})\; ,
\end{equation}
and thus the statistical mixture of coherent states with unknown phase reads:
\begin{equation}
    \rho \approx (1-|\alpha|^2)\ket{0}\bra{0}+|\alpha|^2\ket{1}\bra{1}\; .
\end{equation}
Therefore, one can identify this statistical mixture with an intermittent single photon source providing that $\varepsilon=|\alpha|^2$ is small. The probability of click will then be $p_e=\eta\varepsilon=\eta\bar n $ in either case.

\subsubsection{Efficiency for coherent states}

In order to describe the effect of coherent states on the detector and to compute $\eta$, we now add an input drive of amplitude $\epsilon=\sqrt{\kappa_b}b_{in}$ where $b_{in}$ is the amplitude of the coherent pulse. We start from Eq.~\eqref{eq:drhoqbdt}, where we neglect $\chi_{bb},\chi_{qb},\kappa_q\ll\kappa_\mathrm{nl},\kappa_b$, and take $\Delta=\chi_{qw}$ :
\begin{eqnarray}
\label{eq:drhoqbdt2}
    \frac{d}{dt}\rho_{qb} &=& \kappa_\mathrm{nl}\mathcal{D}[\hat b\hat\sigma^\dag]\rho_{qb} + \kappa_{b}\mathcal{D}[\hat b]\rho_{qb}
    +\epsilon[\hat b-{\hat b}^\dag,\rho_{qb}]\;.
\end{eqnarray}

We would now like to calculate the qubit excited state population $p_e$ as a function of the number of photons in the incoming wave-packet of length $T$: $\bar n_{in}=|b_{in}|^2\times T$, where we assume for simplicity that $b_{in}$ is time-independent. We may write the solution of Eq.~\eqref{eq:drhoqbdt2} in the general form $\rho_{qb}=\rho_{gg}\ket{g}\bra{g}+\rho_{ge}\ket{g}\bra{e}+\rho_{eg}\ket{e}\bra{g}+\rho_{ee}\ket{e}\bra{e}$, and we are interested in $p_e=\tr(\rho_{ee})$. We find
\begin{eqnarray}
\label{eq:drhoggdt}
\frac{d}{dt}\rho_{gg}&=&-\frac{\kappa_\mathrm{nl}}{2}\left(\bb^\dag\bb\rho_{gg}+\rho_{gg}\bb^\dag\bb \right)+\kappa_b\DD[\bb]\rho_{gg}+\epsilon[\bb-\bb^\dag, \rho_{gg}]\\
\label{eq:drhoeedt}
\frac{d}{dt}\rho_{ee}&=&\kappa_\mathrm{nl}\bb\rho_{gg}\bb^\dag+\kappa_b\DD[\bb]\rho_{ee}+\epsilon[\bb-\bb^\dag, \rho_{ee}]
\end{eqnarray}

Hence
\begin{equation}
\label{eq:dpedt}
    \frac{d}{dt}p_e = \kappa_\mathrm{nl}\tr(\bb\rho_{gg}\bb^\dag)\;,
\end{equation}
and we now need to solve Eq.~\eqref{eq:drhoggdt} for which, remarkably, we find a simple ansatz:

\begin{eqnarray}
    \rho_{gg}(t) &=& \exp\left( -4\frac{\epsilon^2}{(\kappa_\mathrm{nl}+\kappa_b)^2}\kappa_\mathrm{nl}t\right)\ket{\beta}\bra{\beta}\\
    \beta&=&-2\epsilon/(\kappa_\mathrm{nl}+\kappa_b)
\end{eqnarray}
where $\ket{\beta}$ is a coherent state of amplitude $\beta$.

Note that $\tr(\bb\rho_{gg}\bb^\dag) = \beta^2\exp\left( -4\frac{\epsilon^2}{(\kappa_\mathrm{nl}+\kappa_b)^2}\kappa_\mathrm{nl}t\right)$, and hence inserting this expression in Eq.~\eqref{eq:dpedt}, we find
\begin{equation}
    \frac{d}{dt}p_e = 4\kappa_\mathrm{nl}\frac{\epsilon^2}{(\kappa_\mathrm{nl}+\kappa_b)^2}\exp\left( -4\frac{\epsilon^2}{(\kappa_\mathrm{nl}+\kappa_b)^2}\kappa_\mathrm{nl}t\right)\;.
\end{equation}
Assuming the qubit in its ground state at $t=0$, we find
\begin{eqnarray}
\label{eq:dpedt2}
p_e(T)&=&1-\exp\left(-\eta|b_{in}|^2T\right)\;,
\end{eqnarray}
where
\begin{eqnarray}
    \eta &=& 4\frac{\kappa_b\kappa_\mathrm{nl}}{(\kappa_\mathrm{nl}+\kappa_b)^2}\;,
\end{eqnarray}
which can be written as 
\begin{eqnarray}
p_e&=&1-\mathbb{P}_0^\eta \underset{\bar n_{in}\to 0}{\sim} \eta\bar n_{in}\;.
\end{eqnarray}

Here $\mathbb{P}_0 = \exp(-\bar n_{in})$ is the population of Fock state 0 of the incoming coherent pulse which has a mean photon number $\bar n_{in}=|b_{in}|^2\times T$ and follows a Poisson distribution. We see from Eq.\eqref{eq:eta} that $\eta = 1$ when $\kappa_\mathrm{nl}=\kappa_b$. 


\subsection{Reset protocol}
For the reset procedure, we switch off the drive on the buffer mode, and instead, we add a resonant drive on the waste port (this drive is sent at $\omega_w-\chi_{qw}$ since the qubit is in $\ket{e}$ when we want to reset). This adds to the Hamiltonian of equation \eqref{eq:Hpp} the following term $H_\text{drive} = \epsilon_w \left(\hat{w}+\hat{w}^\dag\right)$, which can be absorbed by replacing the operator $\bb\hat\sigma^\dag\rightarrow  \bb\hat\sigma^\dag + \epsilon_w/g_3$. Using this modified expression, the rest of the calculation follows. Note that
\begin{equation}
    \kappa_\mathrm{nl}\mathcal{D}[\bb\hat{\sigma}^\dag+\frac{\epsilon_w}{g_3}]\rho_{qb} = \kappa_\mathrm{nl}\mathcal{D}[\bb\hat\sigma^\dag]\rho_{qb} + \kappa_\mathrm{nl}\frac{\epsilon_w}{2g_3}[\bb\hat\sigma^\dag-\bb^\dag\hat\sigma,\rho_{qb}]
\end{equation}
This leads to
\begin{eqnarray}
\label{eq:drhoqbdtreset}
    \frac{d}{dt}\rho_{qb} &=& \kappa_{b}\mathcal{D}[\hat b]\rho_{qb}
    +\kappa_\mathrm{nl}\mathcal{D}[\hat b\hat\sigma^\dag]\rho_{qb}+\epsilon_{nl}[\bb\hat\sigma^\dag-\bb^\dag\hat\sigma,\rho_{qb}]\;,
\end{eqnarray}
where $\epsilon_{nl} = \kappa_\mathrm{nl}\frac{\epsilon_w}{2g_3}$. Instead of repeating the entire analysis of the previous sections, we will proceed using a useful analogy. We see that Eq.~\eqref{eq:drhoqbdtreset} can be mapped to Eq.~\eqref{eq:drhoqbdt2} by making the following substitutions $\kappa_{b}\leftrightarrow\kappa_\mathrm{nl}$, $\bb \leftrightarrow \bb\hat\sigma^\dag$ and $\epsilon_{nl} \leftrightarrow \epsilon$. We have shown that the dynamics of Eq.~\eqref{eq:drhoqbdt2} leads the qubit to dissipate from its ground to its excited state at rate $4\frac{\epsilon^2}{(\kappa_\mathrm{nl}+\kappa_b)^2}\kappa_\mathrm{nl}$ (see Eq.~\eqref{eq:dpedt2}). By analogy, the dynamics of Eq.~\eqref{eq:drhoqbdtreset} leads the qubit to dissipate from its excited state to its ground state at a rate
\begin{eqnarray}
    \kappa_\mathrm{reset} & = & 4\frac{\epsilon_{nl}^2}{(\kappa_{b}+\kappa_\mathrm{nl})^2}\kappa_{b}\\
    &=&4\frac{\epsilon_w^2}{(\kappa_{b}+\kappa_\mathrm{nl})^2}\kappa_\mathrm{nl}\frac{\kappa_b}{\kappa_w}\;.
\end{eqnarray}


\subsection{Spectroscopic characterization of the detector}

\begin{figure}
\includegraphics[width=\columnwidth]{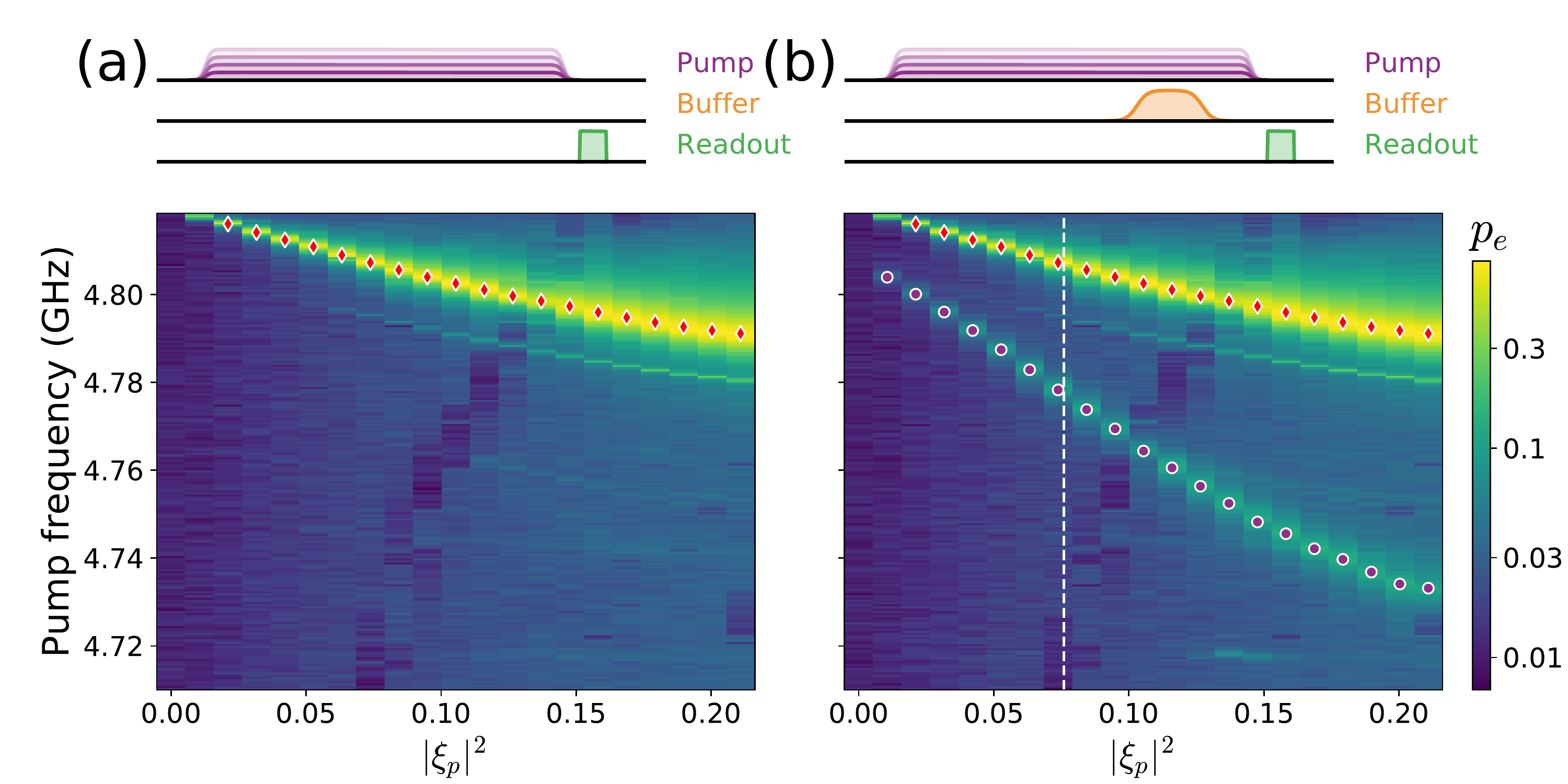}
 \caption{Complete spectroscopic data: (\textbf{a}-\textbf{b}) The excited state population of the qubit $p_e$ (color-coded) is represented as a function of the pump amplitude (x-axis) and pump-frequency (y-axis). (\textbf{a}) When there is no incoming photon on the buffer port, we find one spectroscopic line (red diamonds) for which the qubit is sent into its excited. This corresponds to a higher order amplification process of the form . We also observe that the background $p_e$ is rising approximately linearly with the pump power at all the frequencies which is consistent with the heating of the qubit bath by the pump. (\textbf{b}) When we add a pulse on the buffer port at frequency $\omega_b$, another line (purple circles) appears. This line corresponds the process $\hat b \hat \sigma^\dagger \hat w^\dagger$ and the white dashed line indicates the chosen pump power which maximize the ratio $\eta/\Gamma_{\mathrm{dc}}$. We can also observe a spectroscopic line which goes upward which seems to cool the qubit and thus decreases the efficiency (see dip in $\kappa_{\mathrm{nl}}$ in Fig.~\ref{fig1}a). This was the main reason for limiting the pump power to the chosen value. Also, past this line, the dark count rate increases too much compared to the gain in the efficiency}
\label{fig_spectro}
\end{figure}

When inserting the optimal pump frequency predicted by Eq.~\eqref{eq:drhoqbdt} ($\Delta = \chi_{qw}$) into the frequency matching condition \eqref{eq:freqmatch2}, we find
\begin{equation}
    \omega_p = \left(\omega_q - 2 \chi_{qq} |\xi_p|^2\right) + (\omega_w- \chi_{qw}(1+|\xi_p|^2)) - (\omega_b-\chi_{qb}|\xi_p|^2) \;.
\label{freq_matching_approx}
\end{equation}
This equation is equivalent to Eq. \eqref{eq:freqmatch} with $\bar \omega_q \equiv \omega_q - 2 \chi_{qq} |\xi_p|^2$, $\omega^e_w \equiv \omega_w - \chi_{qw}(1+|\xi_p|^2)$ and $\omega^g_b \equiv \omega_g-\chi_{qb}|\xi_p|^2$. Notice that in the main text, for clarity, we have neglected the AC-Stark shifts on the buffer and waste since $|\xi_p|^2\chi_{qb,w}<|g_3|$. To match this condition, we realize a calibration experiment where the pump power and frequency are scanned over the relevant range of parameter space and we measure the probability $p_e$ to find the qubit in the excited state. We first perform a control experiment, where the buffer is left undriven (Figure \ref{fig_spectro}a). The experiment is then repeated with a small coherent drive pulse on the buffer (Figure \ref{fig_spectro}b). The line corresponding to the process $\hat b \hat \sigma^\dagger \hat w^\dagger$ is clearly identified since the qubit excitation probability $p_e$ vanishes in the absence of buffer drive. For each pump power, the frequency $f_{p0}$ that maximizes the qubit excitation probability within the relevant line is represented with a red dot on Fig. \ref{eq:freqmatch_supp_mat}, and it is also reported in Fig. \ref{fig1}a. The linear dependence of $f_{p0}$ as a function of $|\xi_p|^2$ is used, together with the independently measured value $\chi_{qq}=146 $ MHz, to calibrate $|\xi_p|^2$ in terms of photon number via the relation \eqref{freq_matching_approx}. We attribute the spurious line appearing in both control and calibration experiments to the 6$^\mathrm{th}$ order non-linear process $\xi_p^{3} \hat b^\dagger \sigma^\dagger \hat \sigma_{ef}^\dagger$, where $\hat \sigma_{ef}$ is the lowering operator between the second and first qubit excited states. Indeed, the frequency and slope of the line as a function of pump-power are in good qualitative agreement with the expected value $\omega_\mathrm{p,spurious} = (\bar \omega_q + \bar \omega_q^{ef} + \omega_b)/3$.

\subsection{Tomography of the itinerant transmitted photon}

The non-linear process at the heart of the single photon detector converts the incoming photons (centered around the buffer frequency $\omega_b$) into photons emitted into the waste line at the waste frequency $\omega_w^e$ (the frequency of the waste when the qubit is in the excited state). The field propagating in the waste line is amplified by a near quantum-limited phase-preserving amplifier and demodulated at the frequency $(\omega_w^g+\omega_w^e)/2$. Examples of demodulated time traces, with a sampling rate of 50 MS/s are shown in figure \ref{fig_tomo}c. Besides providing a strongly damped mode for the converted photons, in our experiment, the waste is used to read-out the qubit state. Consequently, the time traces are split in two consecutive segments (see Fig. \ref{fig_tomo}c). The first part of the traces of duration $t_m = 3.4~\mu$s contains an excess noise directly attributable to the up-converted photon and are used to reconstruct the quantum state of propagating field. The second part of the traces of duration $t_r=0.6\,\mu$s, during which the waste cavity is coherently driven by the readout pulse, are used to infer the qubit state.

\subsubsection{Determination of the outgoing photon mode shape}
The complex mode-shape in which single-photons are emitted by the waste corresponds to the input pulse mode-shape which is filtered by the buffer and waste bandwidth as well as the qubit thresholding response, it also undergoes an overall frequency conversion. We employ the method described in Ref. \cite{Morin2013} to determine this waveform experimentally from the 88000 traces where the qubit was detected in state $|e\rangle$: we first combine the raw I and Q quadratures from the digitizer to form the complex discrete-time traces:
\begin{equation}
z(t_j) = I(t_j) + i Q(t_j).
\end{equation}
We then compute the autocorrelation matrix $\langle z^*(t_j) z(t_k) \rangle$ by averaging the two-time correlators over the full dataset. The autocorrelation matrix is diagonalized to find the basis of uncorrelated temporal modes. While most of the eigenvalues have a comparable weight, corresponding to the vacuum and amplifier noise in the unpopulated temporal modes, the largest eigenvalue is significantly larger (by $\sim 25$ \%). The corresponding complex eigenfunction $f(t_j)$ provides thereby the optimal temporal mode, as plotted in Fig. \ref{fig4}c and \ref{fig_tomo}c and is normalized such that $\sum_{j}|f(t_j)|^2 = 1$.

\subsubsection{Photon reconstruction}

\begin{figure}
\includegraphics[width=\columnwidth]{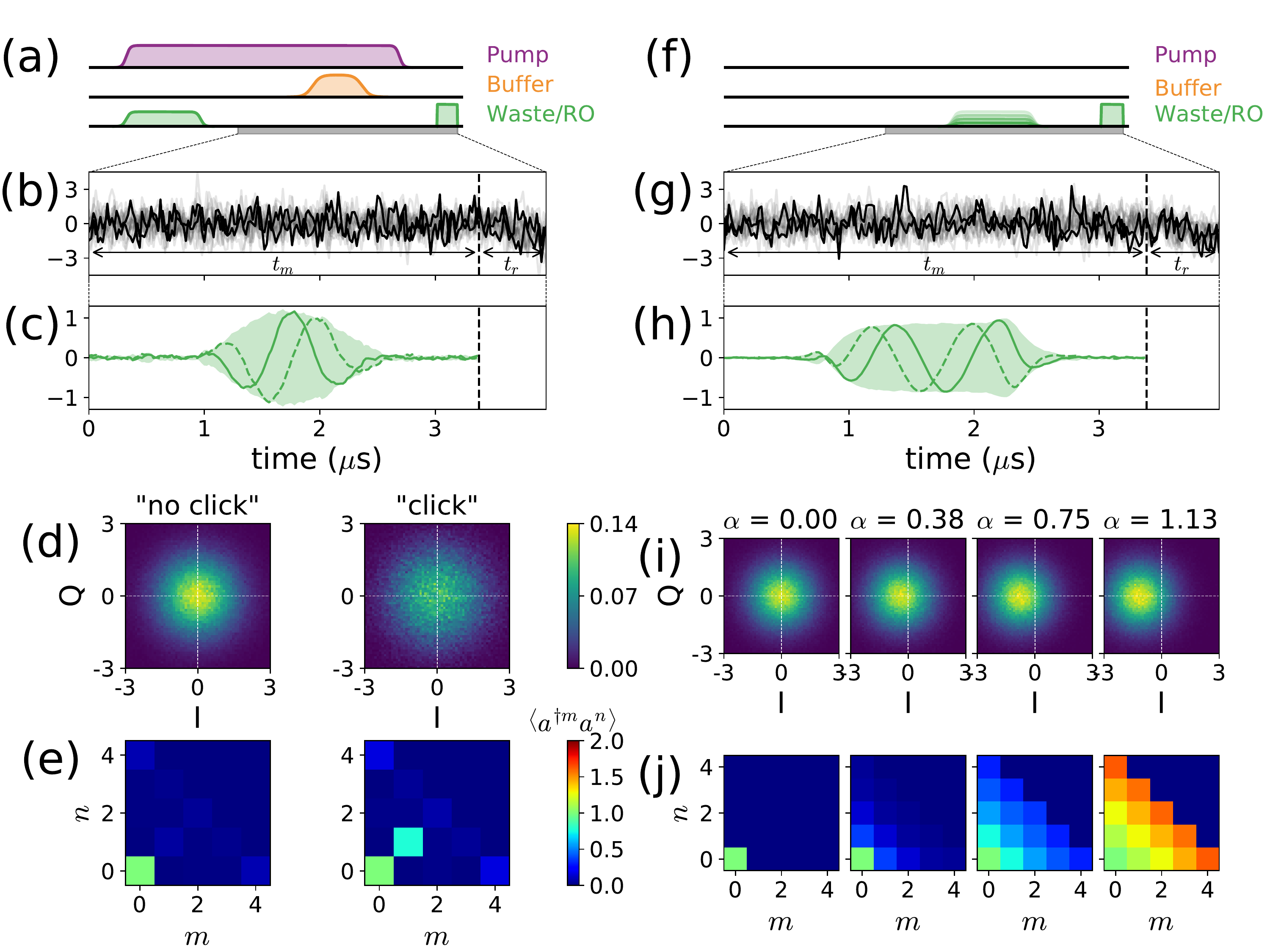}
 \caption{Tomography of the itinerant field states: measurement of the field emitted during a detection sequence (left) and calibration experiment (right).  (a, e) Pulse sequences for the state-reconstruction experiment (a) and amplification calibration (e). The field emitted by the waste is monitored continuously by a phase preserving amplifier (grey rectangle). (b, g) Experimental traces demodulated at $(\omega_w^g + \omega_w^e)/2$ (only one quadrature is represented): the first part of the trace (duration $t_m$) is used for the itinerant field-state reconstruction while the second part (duration $t_r$) is used for qubit state discrimination. (c-h) Mode-shape determined experimentally for the single-photon (c) and coherent fields (h) used for calibration. (d, i) Measured phase space distribution of the propagating field before amplification $\hat S/\sqrt{G}$, it corresponds to the Husimi-Q representation of the quantum state broadened by the equivalent input noise of the amplifier  
 expressed in unit of square-root of photons. The distribution is conditioned on the detection of the qubit in $|g\rangle$ or $|e\rangle$ (d) and for the various coherent state amplitudes used for calibration (i). (e,j) Experimental reconstruction of the moments of the distribution $\langle (\hat a^\dagger )^n \hat a^m \rangle_\rho$ obtained by inverting Eq. \eqref{eq:moments_signal_noise}.}
\label{fig_tomo}
\end{figure}

Now that the temporal-shape of the itinerant mode is identified, we will focus on its quantum state.
Following Ref. \cite{Eichler2011}, we infer the quantum state in the mode-shape $f$ by calculating the various moments of the complex amplitude distribution:
\begin{equation}
S = \sum_j{f^*(t_j) z(t_j)}.
\end{equation}
The experimental phase-space distribution of $S$ conditioned on measuring the qubit in $\ket{g}$ and $\ket{e}$ is presented in Fig.~\ref{fig_tomo}d. The measured distributions --that would correspond to the field's Q-function for a quantum-limited detection-- are significantly broadened by the amplifier's classical-noise. However, a clear excess-noise is visible on the distribution corresponding to $\ket{e}$, that reflects the increased fluctuations in the mode of interest due to the emitted photon.
On a formal level, the observable $\hat S$ can be expressed as a function of the the amplifier gain $G$, the annihilation operator $\hat a$ in mode $f$, and the annihilation operator $\hat h$ in an external mode describing the noise added by the amplifier
\begin{equation}
    \hat S = \sqrt{G} \left(\hat a  + \hat h^\dagger \right).
\end{equation}
Under the assumption that the amplifier noise $\hat h$ is uncorrelated with the signal $\hat a$, we get the following expression for the moments of $\hat S$ \cite{Eichler2011}
\begin{equation}
    \langle ( \hat S^\dagger )^{n} \hat S^m \rangle_{\rho} = G^{(n+m)/2} \sum_{i,j=0}^{n,m}\begin{pmatrix} m \\ j \end{pmatrix}\begin{pmatrix} n\\ i \end{pmatrix} \langle (\hat a^\dagger )^i \hat a^j \rangle_\rho \langle  \hat h^{n-i}(\hat h^\dagger)^{m-j} \rangle.
    \label{eq:moments_signal_noise}
\end{equation}
Remarkably, the moments $\langle\hat h^{n}(\hat h^\dagger)^{m} \rangle = G^{-(n+m)/2} \langle ( \hat S^\dagger )^{n} \hat S^m \rangle_{\ket{0}\bra{0}}$ can be retrieved experimentally by performing a calibration experiment with the mode $\hat a$ in the vacuum state, where $\langle ( \hat S^\dagger )^{n} \hat S^m \rangle_{\ket{0}\bra{0}}$ comes from the raw data and the only remaining unknown is the gain $G$. 

The gain $G$ is calibrated by a control experiment. We send a well calibrated pulse (see \cite{LescannePRApp2019}) on the waste port which reflects back and gets detected by the digitizer (Fig.~\ref{fig_tomo}f,g). Provided the waste is overcoupled to its transmission line (see table \ref{table}), the coherent state in the reflected mode-shape (Fig.~\ref{fig_tomo}h) is equal to the incoming one. Assuming the noise is unbiased, we have $\langle \hat S \rangle = \sqrt{G} \langle \hat a \rangle$ and hence measuring $\langle \hat S \rangle$ for 3 values of $\langle \hat a \rangle$ (Fig.~\ref{fig_tomo}i) we deduce $G$.



The relation \eqref{eq:moments_signal_noise} can then be inverted to obtain the moments of the field's distribution $\langle (\hat a^\dagger )^n \hat a^m \rangle_{\rho_\mathrm{exp}}$ (Fig.~\ref{fig_tomo}e,j). We perform this reconstruction up to $4^\mathrm{th}$ order ($n+m \le 4$). We finally perform a maximum-likelihood estimation of the field's density matrix $\rho_\mathrm{ml}$ by minimizing the distance between the measured and expected moments:
\begin{equation}
    \mathcal{D}(\rho_\mathrm{exp}, \rho_\mathrm{ml})=\sum_{\substack{m,n \\m+n\le4}}\left| \langle (\hat a^\dagger )^n \hat a^m \rangle_{\rho_\mathrm{exp}} - \langle (\hat a^\dagger )^n \hat a^m \rangle_{\rho_\mathrm{ml}} \right|^2.
\end{equation}
The Wigner-functions plotted in Figure \ref{fig4} are calculated from the density matrix $\rho_\mathrm{ml}$.

\begin{acknowledgements}
\end{acknowledgements}





\end{document}